\input harvmac
\input graphicx

\def\Title#1#2{\rightline{#1}\ifx\answ\bigans\nopagenumbers\pageno0\vskip1in
\else\pageno1\vskip.8in\fi \centerline{\titlefont #2}\vskip .5in}
%

%
%
\ifx\includegraphics\UnDeFiNeD\message{(NO graphicx.tex, FIGURES WILL BE IGNORED)}
\def\figin#1{\vskip2in}
\else\message{(FIGURES WILL BE INCLUDED)}\def\figin#1{#1}
\fi
\def\Fig#1{Fig.~\the\figno\xdef#1{Fig.~\the\figno}\global\advance\figno
 by1}
%
%
%
%
\def\Ifig#1#2#3#4{
\goodbreak\midinsert
\figin{\centerline{
\includegraphics[width=#4truein]{#3}}}
\narrower\narrower\noindent{\footnotefont
{\bf #1:}  #2\par}
\endinsert
}
%
%
\font\ticp=cmcsc10

\def\hf{{1\over 2}}

\def\calo{{\cal O}}
\def\calh{{\cal H}}
\def\hbh{\calh_{BH}}
\def\hnear{\calh_{near}}
\def\hfar{\calh_{far}}

\def\roughly#1{\mathrel{\raise.3ex\hbox{$#1$\kern-.75em\lower1ex\hbox{$\sim$}}}}
\def\calN{{\cal N}}

\def\ahat{{\hat a}}

\def\uhat{{\hat u}}
\def\tret{t_{\rm xfer}}
\def\Ebar{{\bar {\cal E}}}
\def\Ibar{{\bar I}}
\def\Jbar{{\bar J}}
\overfullrule=0pt
%
%

\lref\Hawkrad{
  S.~W.~Hawking,
  ``Particle Creation By Black Holes,''
  Commun.\ Math.\ Phys.\  {\bf 43}, 199 (1975)
  [Erratum-ibid.\  {\bf 46}, 206 (1976)].
}
\lref\BHMR{
  S.~B.~Giddings,
  ``Black holes and massive remnants,''
Phys.\ Rev.\  {\bf D46}, 1347-1352 (1992).
[hep-th/9203059].
}
\lref\thooholo{
  G.~'t Hooft,
  ``Dimensional reduction in quantum gravity,''
  arXiv:gr-qc/9310026.
}
\lref\sussholo{
  L.~Susskind,
  ``The World As A Hologram,''
  J.\ Math.\ Phys.\  {\bf 36}, 6377 (1995)
  [arXiv:hep-th/9409089].
}
\lref\AMPS{
  A.~Almheiri, D.~Marolf, J.~Polchinski and J.~Sully,
  ``Black Holes: Complementarity or Firewalls?,''
[arXiv:1207.3123 [hep-th]].
}
\lref\Mathinfall{
  S.~D.~Mathur and C.~J.~Plumberg,
 ``Correlations in Hawking radiation and the infall problem,''
JHEP {\bf 1109}, 093 (2011).
[arXiv:1101.4899 [hep-th]].
}
\lref\SBGmodels{
  S.~B.~Giddings,
  ``Models for unitary black hole disintegration,''
Phys.\ Rev.\ D {\bf 85}, 044038 (2012).
[arXiv:1108.2015 [hep-th]].
}
\lref\BaFitherm{
  T.~Banks, W.~Fischler,
  ``Space-like Singularities and Thermalization,''
[hep-th/0606260].
}
\lref\GiddingsUE{
  S.~B.~Giddings,
  ``Black holes, quantum information, and unitary evolution,''
Phys.\ Rev.\ D {\bf 85}, 124063 (2012).
[arXiv:1201.1037 [hep-th]].
}
\lref\Compl{
  L.~Susskind, L.~Thorlacius and J.~Uglum,
  ``The Stretched horizon and black hole complementarity,''
Phys.\ Rev.\ D {\bf 48}, 3743 (1993).
[hep-th/9306069].
}
\lref\Lenny{
  L.~Susskind,
  ``The Transfer of Entanglement: The Case for Firewalls,''
[arXiv:1210.2098 [hep-th]].
}
\lref\Raphael{
  R.~Bousso,
  ``Complementarity Is Not Enough,''
[arXiv:1207.5192 [hep-th]].
}
\lref\LoTh{
  K.~Larjo, D.~A.~Lowe and L.~Thorlacius,
  ``Black holes without firewalls,''
[arXiv:1211.4620 [hep-th]].
}
\lref\BaFi{
  T.~Banks,
  ``Holographic Space-Time: The Takeaway,''
[arXiv:1109.2435 [hep-th]].
}
\lref\LawrenceSG{
  A.~E.~Lawrence and E.~J.~Martinec,
  ``Black hole evaporation along macroscopic strings,''
Phys.\ Rev.\ D {\bf 50}, 2680 (1994).
[hep-th/9312127].
}
\lref\BrownUN{
  A.~R.~Brown,
 ``Tensile Strength and the Mining of Black Holes,''
[arXiv:1207.3342 [gr-qc]].
}
\lref\algquant{
  R.~Haag,
  {\sl Local quantum physics: Fields, particles, algebras,}
Berlin, Germany: Springer (1992) 356 p. (Texts and monographs in physics).
}
\lref\NLvsC{
  S.~B.~Giddings,
  ``Nonlocality versus complementarity: A Conservative approach to the information problem,''
Class.\ Quant.\ Grav.\  {\bf 28}, 025002 (2011).
[arXiv:0911.3395 [hep-th]].
}
\lref\GiNe{
  S.~B.~Giddings and W.~M.~Nelson,
 ``Quantum emission from two-dimensional black holes,''
Phys.\ Rev.\ D {\bf 46}, 2486 (1992).
[hep-th/9204072].
}
\lref\Fisheretal{M.P.A.~Fisher, P.B.~Weichman, G.~Grinstein, and D.S.~Fisher, ``Boson localization and the superfluid-insulator transition," Phys.\ Rev.\ B {\bf 40}, 546 (1989).}
\lref\Mathurbit{
  S.~D.~Mathur,
  ``The Information paradox: A pedagogical introduction,''
Class.\ Quant.\ Grav.\  {\bf 26}, 224001 (2009).
[arXiv:0909.1038 [hep-th]].
}
\lref\LPSTU{
  D.~A.~Lowe, J.~Polchinski, L.~Susskind, L.~Thorlacius and J.~Uglum,
  ``Black hole complementarity versus locality,''
  Phys.\ Rev.\  D {\bf 52}, 6997 (1995)
  [arXiv:hep-th/9506138].
}
\lref\Page{
  D.~N.~Page,
  ``Average entropy of a subsystem,''
Phys.\ Rev.\ Lett.\  {\bf 71}, 1291 (1993).
[gr-qc/9305007];
 ``Information in black hole radiation,''
  Phys.\ Rev.\ Lett.\  {\bf 71}, 3743 (1993)
  [arXiv:hep-th/9306083].
}
\lref\GiSh{
  S.~B.~Giddings and Y.~Shi,
 ``Quantum information transfer and models for black hole mechanics,''
[arXiv:1205.4732 [hep-th]].
}
\lref\QBHB{
  S.~B.~Giddings,
  ``Quantization in black hole backgrounds,''
  Phys.\ Rev.\  D {\bf 76}, 064027 (2007)
  [arXiv:hep-th/0703116].
}
\lref\bigboosts{
  S.~B.~Giddings,
 ``Black hole information, unitarity, and nonlocality,''
Phys.\ Rev.\ D {\bf 74}, 106005 (2006).
[hep-th/0605196];
``(Non)perturbative gravity, nonlocality, and nice slices,''
Phys.\ Rev.\ D {\bf 74}, 106009 (2006).
[hep-th/0606146].
}
\lref\ParkerJM{
  L.~Parker,
  ``Probability Distribution of Particles Created by a Black Hole,''
Phys.\ Rev.\ D {\bf 12}, 1519 (1975).
}
\lref\UnWamine{
  W.~G.~Unruh and R.~M.~Wald,
  ``Acceleration Radiation and Generalized Second Law of Thermodynamics,''
Phys.\ Rev.\ D {\bf 25}, 942 (1982)\semi
``How to mine energy from a black hole," Gen. Rel. and Grav. {\bf 15}, 195 (1983).
}
\lref\Frolov{
  V.~P.~Frolov and D.~Fursaev,
  ``Mining energy from a black hole by strings,''
Phys.\ Rev.\ D {\bf 63}, 124010 (2001).
[hep-th/0012260]\semi
  V.~P.~Frolov,
 ``Cosmic strings and energy mining from black holes,''
Int.\ J.\ Mod.\ Phys.\ A {\bf 17}, 2673 (2002).
}
\lref\fuzz{
  S.~D.~Mathur,
  ``Fuzzballs and the information paradox: A Summary and conjectures,''
[arXiv:0810.4525 [hep-th]].
}
\lref\Arkank{N. Arkani-Hamed, talk at the KITP conference {\sl String phenomenology 2006};  N.~Arkani-Hamed, S.~Dubovsky, A.~Nicolis, E.~Trincherini and G.~Villadoro,
  ``A Measure of de Sitter entropy and eternal inflation,''
JHEP {\bf 0705}, 055 (2007).
[arXiv:0704.1814 [hep-th]].
}
\lref\KlebanovEH{
  I.~R.~Klebanov, L.~Susskind and T.~Banks,
  ``Wormholes And The Cosmological Constant,''
Nucl.\ Phys.\ B {\bf 317}, 665 (1989)..
}
\Title{
\vbox{\baselineskip12pt
}}
{\vbox{\centerline{Nonviolent nonlocality}
}}
\centerline{{\ticp 
Steven B. Giddings\footnote{$^\ast$}{Email address: giddings@physics.ucsb.edu}  
} }
\centerline{\sl Department of Physics}
\centerline{\sl University of California}
\centerline{\sl Santa Barbara, CA 93106}
\vskip.40in
\centerline{\bf Abstract}
If quantum mechanics governs nature, black holes must evolve unitarily, providing  a powerful constraint on the dynamics of quantum gravity.  Such evolution apparently must in particular be nonlocal, when described from the usual semiclassical geometric picture, in order  to transfer quantum information into the outgoing state.  While such transfer from a disintegrating black hole has the dangerous potential to be violent to generic infalling observers, this paper proposes the existence of a more innocuous form of information transfer, to relatively soft modes in the black hole atmosphere.  Simplified models for such nonlocal transfer are described and parameterized, within a possibly more basic framework of networked Hilbert spaces.  Sufficiently sensitive measurements by infalling observers may detect departures from Hawking's predictions, and in generic models black holes decay more rapidly.  Constraints of consistency -- internally and with known and expected features of physics -- restrict the form of information transfer, and should provide important guides to discovery of the principles and mechanisms of the more fundamental nonlocal mechanics.

\vskip.3in
\Date{}

\newsec{Introduction}

If black hole evolution is unitary, as many now expect, apparently Hawking evaporation\Hawkrad\ must receive some nonlocal modification; early proposals for its form appear in \refs{\BHMR,\thooholo,\sussholo}.  A particularly important question is whether such nonlocal physics is violent enough to produce drastic departure from the semiclassical black hole geometry.  This paper explores a conjecture of {\it nonviolent nonlocality}:
the full quantum description of black holes matches the semiclassical picture of black holes, to a good approximation, and in particular permits infalling observers to cross the semiclassical horizon.  Modifications of Hawking radiation may be detectable, but not violent.

More specifically, we assume that the fundamental dynamics is formulated within a sufficiently general framework for quantum mechanics.  We explore the possibility that the local quantum field theory (LQFT) description of black hole physics, quantizing fluctuations about a backreaction-corrected semiclassical geometry, is an approximation to this that is useful for describing many features of black hole evolution, and in particular the possibility of observers entering a semiclassical black hole interior.  

This possibility contrasts with other scenarios that have been explored.  For example, the black hole might evolve into an object where a surface or interface replaces the horizon, and such a surface is expected to be damaging to infalling observers.  Falling onto a neutron star provides an analogy, but an exotic ``star-like" object replacing a black hole is expected to be even more compact, with size that could be comparable to the Schwarzschild radius.
We will refer to such an object as a ``massive remnant;" this type of scenario was first explored in \BHMR.  More recently, ref.~\refs{\AMPS} has argued that unitary evolution requires a ``firewall" of highly energetic quanta at or near the would-be horizon, providing one way of realizing such a remnant interface.   The fuzzball scenario\refs{\fuzz} appears to be in the same category; the horizon is replaced by stringy states, carrying the quantum information of the would-be black hole.  While it has been proposed\refs{\Mathinfall} that an infalling observer can pass painlessly through the fuzzball near-horizon interface, into a ``dual" image, this picture seems difficult to realize in detail, and one might instead expect a destructive impact from interaction with microstructure at the interface. 

Such scenarios are  violent not only to infalling observers, but also to the known laws of physics.  They require nonlocal evolution, if the interface is to emerge from within an initial black hole, and moreover evolution into a state that is a drastic departure from our usual picture of a black hole.  They thus manifest violent nonlocality.
 Indeed, depending on the formation time for such a remnant, we might need to describe the would-be black hole at the center of our galaxy as a very different sort of object.

While nonlocality is apparently required, we should ask whether a less violent version is allowed, without implying any deeper inconsistency\NLvsC.  That this is a non-trivial question is illustrated by the arguments of \AMPS, which provide constraints on  such a picture.   
Indeed, in the models of \refs{\SBGmodels,\GiddingsUE}, it was recognized that imprinting quantum information on outgoing quanta is a potentially dangerous modification to the Hawking state; if such quanta are propagated backwards, without modification to LQFT rules, they become highly blueshifted at the horizon.  Also, arguments of \AMPS\ appear to indicate that the alternate scenario of ``black hole complementarity\refs{\Compl}"\foot{Here, we have in mind complementarity of variables in the sense of Bohr, as described by {\it e.g.} \Lenny.} instead yields a firewall\refs{\Lenny,\Raphael}.\foot{Though for counterpoints, see \refs{\LoTh}.}

Once nonlocality, at least with respect to the usual LQFT/semiclassical picture, is under consideration, we can explore its realizations.  In particular, \refs{\SBGmodels,\GiddingsUE} suggested giving up postulate II of \refs{\Compl}, which states that evolution outside the horizon is that of LQFT, providing a possible way to evade the problem of singular behavior at the horizon, that was later elaborated in \AMPS.  The present paper will extend and provide further detail to a possible realization of such dynamics.  One goal is to try to realize the proposed general picture of nonviolent nonlocality, in a way that is not manifestly inconsistent with other principles of physics.  More detail should help us explore and refine conditions both of internal consistency, and of consistency with expected or known aspects of physics.   

If such a picture is correct, its development could provide a guide to the physical mechanics and principles of quantum gravity, perhaps like the initially ad hoc rules for the quantum atom helped guide development of quantum mechanics.  The need to describe unitary black hole evolution is an important constraint on physics.   While the ultimate formulation of quantum gravity may well be a large departure from LQFT, we initially attempt to parameterize that departure in the language of LQFT.  Such a parameterization may be fundamentally wrong, but useful nonetheless, just as quantum rules were initially presented in incorrect classical language.  One motivation for this is that the more complete picture should match onto LQFT and its experimentally-tested features through correspondence.   The corrections may also predict measurable effects, given sufficiently careful measurements -- certainly of the details of black hole radiation, but also possibly of more coarse-grained features, such as the radiated power from a black hole.  Following  \refs{\NLvsC,\SBGmodels,\GiddingsUE}, this paper proposes that this is possible without necessarily destroying infalling observers.  

A possible fundamental framework, extending beyond LQFT, is that of networked Hilbert spaces, where states and quantum information, together with an overlapping structure of tensor factors, give a structure that is more basic than spacetime.  This was proposed in \GiddingsUE; a prior proposal with some similarities (and important differences) is that of ``holographic spacetime\refs{\BaFi}."   Indeed, LQFT may be cast in a similar framework\refs{\algquant}.

\newsec{Evolution in networked Hilbert spaces}

\subsec{LQFT description}
	
Our goal is to describe quantum information transfer from a black hole that is required to yield unitary evolution.  While locality indicates this is forbidden in LQFT, we expect that LQFT does accurately describe many features of the black hole and its state.  For that reason, while the {\it fundamental} description of black hole mechanics is apparently not via LQFT expanded about a semiclassical background spacetime, we will attempt to model and parameterize the necessary evolution, at least approximately, as a departure from LQFT.

 \Ifig{\Fig\slices}{Shown are representative slices from three types of slicings of a black hole spacetime.   In the static approximation, the family of slices is generated by translating the representative by Schwarzschild time translations; small corrections to this give the slicing in the presence of black hole decay.}{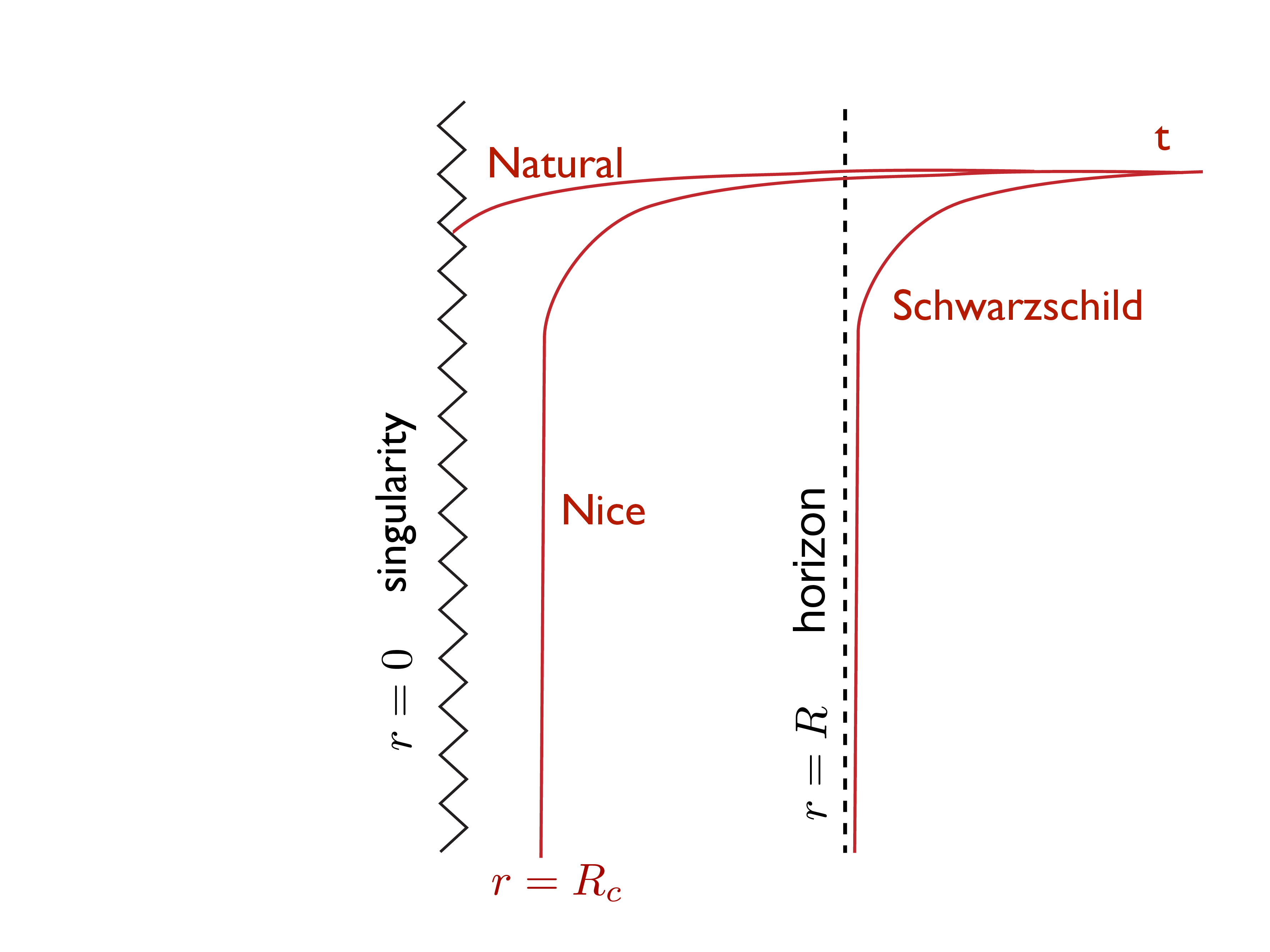}{5}

We begin by reviewing a LQFT description of black hole evolution, and then will study its modification.  Consider time evolution in the canonical formalism, with respect to a particular time slicing of the semiclassical spacetime.  Three commonly-used types of slices are shown in \slices.  The first were called {\it natural} slices in \refs{\NLvsC}; time slices defined by initially synchronized clocks on a constellation of infalling satellites will have this character, and in particular terminate in the singularity.  The second are {\it nice slices}, initially discussed in \LPSTU.  In the static Schwarzschild solution, the last are {\it Schwarzschild slices}.  We take each kind of slice to be asymptotic to such Schwarzschild slices, and so label them by Schwarzschild time $t$ at infinity.

For some purposes we expect the static approximation to be good.
For an evaporating black hole, 
\eqn\evapfrac{{R\over M} {dM\over dt} \sim {1\over S_{BH}}\ ,}
where $R$ is the Schwarzschild radius and $S_{BH}$ is the Bekenstein-Hawking entropy, so the fractional change in mass in time $R$ is tiny; this holds even if the outward energy flux significantly exceeds the Hawking rate, {\it e.g.} to accommodate needed information flow.
This suggests that to a good approximation one can neglect shrinkage of the black hole, over intermediate timescales.  In the limit where we do, the full family of slices can be constructed by simply translating one of the slices in \slices\ forward in time.  Explicit parameterizations of each of the three kinds of slicings, in a static geometry, are given in \GiddingsUE.  Similar families of slices can be constructed outside the static limit, by appropriate adjustments of the slice parameters with time.

The metric in such a slicing can be written in ADM form,
\eqn\ADMmet{ds^2 = -N^2 dt^2 + g_{ij} (dx^i+N^i dt) (dx^j+N^j dt)\ ,}
and the LQFT evolution operator takes the form (in $D$ spacetime dimensions)
\eqn\ulqft{U_{LQFT}=\exp\left\{ -i \int dt d^{D-1} x\sqrt{{}^{D-1}g}  N{\cal H}\right\}\ }
where $\cal H$ is the hamiltonian density.  In the present discussion we view spin as a largely inessential complication, and so take the example of a scalar field $\phi$, with canonical momentum
\eqn\pidef{ \pi = {\partial_t \phi - N^i \partial_i\phi\over N}= n^\mu\partial_\mu\phi\quad ;\quad [\pi(x,t),\phi(y,t)] = -i{\delta^{D-1}(x-y)\over \sqrt{{}^{D-1}g}}}
and hamiltonian\foot{Note that this definition differs from some ADM treatments by combining both constraint terms, multiplied by $N$, $N^i$ respectively.}
\eqn\Hphi{N{\cal H} = \hf N(\pi^2 + g^{ij}\partial_i \phi \partial_j \phi )+ N^i \pi\partial_i\phi\ .}
It is also useful to expand the field $\phi$ in a basis of modes $U_I$, 
\eqn\phiexp{\phi(x,t)=\sum_I \left[A_I U_I(x,t) + A_I^\dagger U_I^*(x,t)\right]  }
that are solutions of the classical wave equation.  With modes normalized in the Klein-Gordon inner product, the $A_I$ obey the usual canonical commutation relations $[A_I,A_J^\dagger]=\delta_{IJ}$.

A decomposition of the Hilbert space into smaller Hilbert spaces of a tensor network, corresponding to the black hole interior, $\calh_{BH}$, the black hole atmosphere, $\calh_{near}$, and the asymptotic region, $\calh_{far}$, can be given through specification of an appropriate set of modes.  For example, working at time $t=0$, a basis $U_I(x,0)$ can be chosen dividing the modes into those localized inside and outside the black hole, and also in the near and far regions.\foot{Of course for the Schwarzschild slicing, the first set is absent.}  One way to do this is to use $U_I$ corresponding to wavepacket solutions, {\it e.g.} the windowed Fourier transformations described in \refs{\Hawkrad,\GiNe,\GiddingsUE}.  This is very explicit in two dimensions, where reflection/gray body factors are not relevant, simplifying dynamics\GiNe.  In a given physical process, at a given time, only modes longer than some shortest wavelength $l_0$ are excited; shorter wavelength modes may be taken to be in the local vacuum.  If an IR cutoff is also present, this makes the Hilbert spaces finite dimensional.

One way to describe the dynamics is in the Schrodinger picture, expanding the field in  constant $U_I(x)=U_I(0,x)$.  Inserting \phiexp\ and the corresponding expression for the momentum into  \Hphi\ then reexpresses the evolution operator as
\eqn\Usch{U_{LQFT}=\exp\{-itH\}\ ,}
where
\eqn\Hexp{H=\sum_{IJ} A^\dagger_I H_{\Ibar J}A_J + A_I^\dagger H_{\Ibar \Jbar}A_J^\dagger +  A_I H_{IJ}A_J + {\rm const.}\ .}
This hamiltonian, which bears some similarity to a non-interacting Bose-Hubbard hamiltonian\refs{\Fisheretal}, describes both ``hopping" of modes from region to region, and production of new excitations, as predicted by Hawking.  Such a description can likewise be given for fields with higher spin.  Particularly simple coarse-grained idealizations of such evolution are the qubit models of \refs{\Mathurbit}, where qubits evolve by hopping, and are produced in the atmosphere of the black hole.

Note two important features.  First, LQFT evolution at the black hole horizon is not time-reversal invariant: the matrix elements $\sim A_{near} A_{BH}^\dagger$, $ A^\dagger_{near} A_{BH}$ preferentially hop modes into the black hole, and, modulo tails of wavepackets, excitations are not allowed to escape from behind the horizon.  Indeed, locality is encoded in $H$ in the statement that it does not propagate excitations outside the lightcone.  (Symmetries and conservation laws must likewise be encoded in these coefficients.)  Second, Schwarzschild and nice slicings in principle give a complete description of the Hilbert space, in the static approximation, albeit with important degeneracies.  But, for a complete description in a natural slicing, one must augment the description with states at the would-be singularity, which we will call the ``core" of the black hole.

\subsec{Information transfer:  modeling nonlocal mechanics}

\Ifig{\Fig\basicp}{Schematic picture of quantum information transfer into and out of a black hole.  Information from infalling matter or early Hawking excitations resides within the black hole, until a time $\tret$ when it transfers into modes in the black hole atmosphere.  At intervening times it may undergo scrambling.  The ``active" atmosphere modes receiving the information may have wavelength much longer than the cutoff, $\sim l_{Pl}$, and up to of order the black hole size $R$.  This can soften effects on infalling observers.  Also shown are representatives of a natural slicing, corresponding to a gauge choice for describing evolution.}{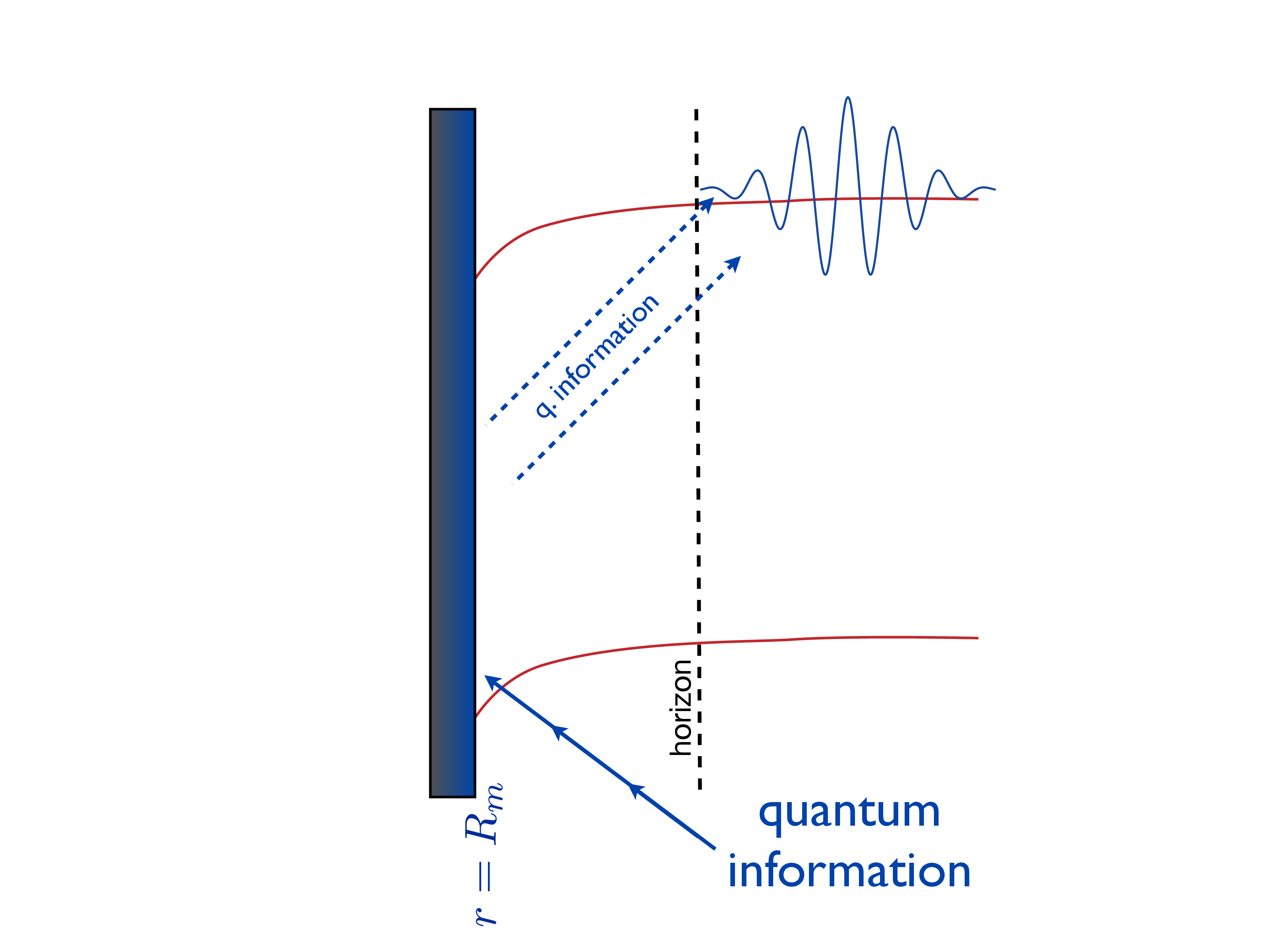}{5}

We next model information leakage from a black hole.  A schematic picture is given in  \basicp.  However, we again stress that such a spacetime picture might be viewed as {\it only an approximation} to a more fundamental underlying mechanics.  Nonetheless, we attempt to approximately parameterize such mechanics in terms of the LQFT picture.  

Suppose that a black hole is formed from a pure state, {\it e.g.} in a high-energy collision.  It then Hawking radiates, and this process builds up entanglement between the outgoing radiation and the internal black hole states.  This entanglement can be characterized  by the von Neumann entropy $S$ of the outgoing radiation.  Next, suppose that at some time $\tret$ information transfer from the black hole begins.  We wish to describe this process.  Arguments given by Page\refs{\Page} indicate that this must happen by the time $t_{Page}\sim R S_{BH}$ where the number of excited degrees of freedom inside the black hole is comparable to the number of Hawking quanta, but we also consider $\tret\ll t_{Page}$.

By information transfer we mean subsystem transfer\refs{\GiSh}, which can be thought of as transfer of degrees of freedom,  or more generally 
nonminimal transfer creating additional entanglement.  Sometimes the former is loosely called ``qubit transfer."  As described in \GiSh, such transfer implies transfer of the entanglement between the early radiation and black hole to entanglement between the early and late radiation.

At the time $\tret$ when the information transfer becomes an important effect, we assume the black hole to still be shrinking slowly, in the sense that the left hand side of \evapfrac\ is small and the static background geometry provides an approximate description.
To model the transfer, we give more detail of the decomposition $\calh = \hbh\otimes\hnear\otimes\hfar$.  We divide the set of modes $U_I$ into a set $\uhat_i$ inside the black hole, a set $u_a$ describing the atmosphere $\hnear$ of modes localized to, say, $r<5R$, and a set $u_\alpha$ of asymptotic modes corresponding to $\hfar$.  Ladder operators $A_I$ are likewise segregated as $\{\ahat_i, a_a, a_\alpha\}$.

\Ifig{\Fig\sliceext}{In one realization of a natural slicing ending at $r=R_m$, one may choose slices that asymptote to nice slices at $r=R_c<R_m$.  This may be used as a bookkeeping device for the ``core" states of the black hole, which may be described by their appearance on the nice slice at $r=R_c$.  In the LQFT description, evolution freezes at $r=R_c$. }{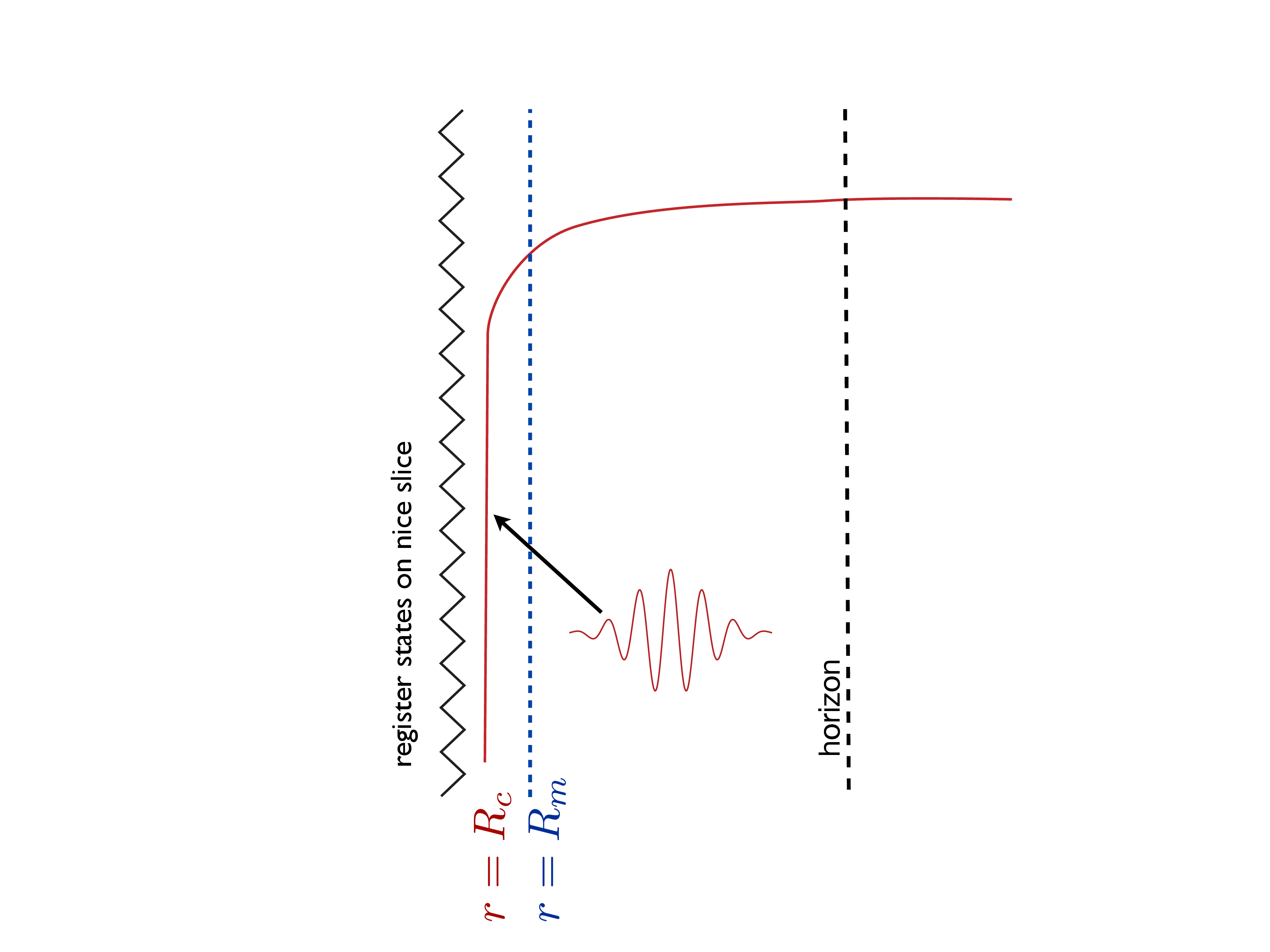}{5}

We consider a  description based on a natural slicing; the next section discusses aspects of gauge transformations between slicings.  Specifically, let us cut off the slice at $r=R_m>0$ to avoid infinite curvature.  Evolution of modes for  $r>R_m$ is essentially that of LQFT.  These modes 
must be augmented by quantum variables describing the states of the black hole core, to give a complete unitary description, tracking the information that falls into $r<R_m$.  In fact, purely as a bookkeeping device, we can consider using nice slices matching our natural slicing for $r>R_m$, but then asymptoting to constant $r=R_c<R_m$, as shown in \sliceext.  LQFT evolution of degrees of freedom then ``freezes" at $r=R_c$, as seen\refs{\QBHB} by vanishing of the lapse $N$ in \ulqft.  While it appears that such frozen evolution cannot be correct over very large times, we can use such a description for our bookkeeping of states, and specifically describe the ``core" states as created by operators $\ahat^\dagger_i$ corresponding to modes within $r<R_m$ in a nice-slice description.  A simpler ``toy" qubit model for these states is described in the appendix.

In order to restore unitarity to black hole evolution, our new dynamics must transfer information from these core states to outgoing degrees of freedom; we also expect it could lead to nontrivial evolution (or scrambling) of these states, prior to transfer.

We can model the former effect by augmenting the hamiltonian with an extra term of the form
\eqn\hnl{H_{NL} = {1\over R} \sum_{i,a} \calN_{{\bar a}i}(t) a^\dagger_a \ahat_i+ {\rm h.c}\ ,}
where the $\calN_{{\bar a}i}$ are dimensionless amplitudes.\foot{In the Schr\"odinger picture, we expect these to be nearly time independent.}  
Generic nonvanishing  
$\calN_{{\bar a}i}$  violate locality, with the detailed form of the violation characterized by the pattern of nonvanishing coefficients.
 Note that in order to be completely explicit, the operators in \hnl\ can be constructed from inner products of wavepackets with the field $\phi$.

In particular, violent vs. nonviolent transfer is governed by which atmosphere modes $u_a$ receive the information.\foot{Clearly one could generalize to transfer to modes of $\hfar$, though this appears less motivated; we might expect the nonlocal information transfer to only be present near the black hole.  Indeed, note also the parallel with bilocal operators induced by spacetime wormholes\KlebanovEH.  In the wormhole case, integration over spacetime enforces energy/momentum conservation for each operator independently, but the background of the present case offers a way to evade such constraints.  Multilocal interactions may also be considered.}  Transfer into a mode $u_a$ that is ultraplanckian at the time of transfer presents a dramatic signal for an infalling observer, and population of enough such modes would be fatal.  But, transfer into longer wavelength modes -- even modes with wavelength much less than $R$ -- can be far less dramatic.  Indeed, if Earth were right now falling into the black hole at the center of our  galaxy, and a modest number of graviton or even photon modes with wavelength $\ll R_{\rm Sgr\,A^*}\sim 15\times 10^6\, {\rm km}$ were populated by such an effect, none of our present detectors would see these excitations.

In the time $t$ after a black hole is created through collapse or collision, of order $t/R$ Hawking particles are emitted, and their partner excitations enter the core region and increase the number of excitations created by the ${\hat a}_i^\dagger$.  Indeed, following this process for the characteristic evaporation time $t\sim R S_{BH}$ is one way to account for the $\calo(\exp\{S_{BH}\})$ internal states of a black hole\refs{\ParkerJM,\GiNe,\QBHB}.  

As noted, in a nice-slice description these excitations are frozen, but after sufficiently long time we expect this description to be inaccurate\refs{\QBHB}.  Possible mixing of these excitations can be modeled by adding to the LQFT hamiltonian, which vanishes on core states in such a description, additional terms of the form
\eqn\hscram{H_{sc}= {1\over R} \sum_{i,j} {\cal S}_{{\bar \imath }j} \ahat_i^\dagger \ahat _j \ ,}
or further terms mixing $\ahat_i\ahat_j$ or $\ahat_i^\dagger\ahat_j^\dagger$.  This hamiltonian  represents {\it scrambling} of the black hole internal states.  As stressed in \refs{\GiddingsUE, \GiSh}, both the transfer of \hnl\ and scrambling of \hscram\ can play roles in quantum information escaping the black hole.

In principle the timescales for transfer and scrambling to operate can be different\refs{\GiddingsUE}.  While internal scrambling may be expected -- and has even been suggested\refs{\BaFitherm}  to arise on a time scale $\sim R$ -- it is not strictly speaking necessary.  This is illustrated in the simplest qubit models of \refs{\SBGmodels,\GiddingsUE}, where qubits are simply transferred out of the black hole interior when they reach the far end of the qubit string; no qubits need be scrambled.  As noted, we need the time $\tret$ where transfer becomes important to satisfy $\tret\roughly< t_{Page}$.  By this time, we need a minimum transfer rate corresponding to $\calo(1)$ qubit per time $R$.  This, however, is a slow leakage rate -- we might think of it as similar to slow escape of helium through the imperfect barrier of a balloon.
If there are $\sim S_{BH}$ internal modes excited, this can be accomplished with amplitudes of generic size\foot{This simple estimate comes from the statement that  $S_{BH}$ degrees of freedom each having a probability $1/S_{BH}$ to transfer per time $R$ leads to ${\cal O}(1)$ degree of freedom transferred in that time.}  $\calN_{{\bar a} i}\sim \calo(1/\sqrt {S_{BH}})$.

We noted the important feature of violation of manifest time-reversal invariance in \Hexp, related to the temporal directionality of the background and slicing.  Such irreversibility could also be a feature of \hnl, in particular allowing one to avoid or suppress the reverse process of quanta in the atmosphere nonlocally propagating into the black hole.  Here the dynamics would differ from leakage through a static barrier.  Of course, ultimately the full dynamics should respect constraints of unitarity, and so reversibility.  However, that is not necessarily manifestly exhibited in the present description of perturbations about a black hole background.

While special models are given in  \refs{\SBGmodels,\GiddingsUE} where the outward energy flux matches that of Hawking, the generic modification of Hawking evolution implies {\it extra flux} of energy.  This can occur both because of higher transfer rates from \hnl\ than the minimal rate, and if transfer is to modes harder than those of energy $\sim 1/R$.  Consistency or other considerations may require information discharge of a black hole to begin at  a time $\tret \ll t_{Page}$.  At such a time, a measure of the internal information is the number of Hawking excitations created, $S\sim \tret/R$, determining the range of $i$ for which excitation is present.

\subsec{Threaded strings, mining, and information overfilling}

Another motivation for the idea of information flux exceeding the natural ``Hawking" rate, $\sim dS_{BH}/dt$ after time $t_{Page}$, comes from possible enhancement of the black hole decay rate either from threading it with a string ({\it e.g.} cosmic) or brane, or from mining\refs{\UnWamine}. In the former case, the string or brane provides extra modes for Hawking radiation, which may have little gray-body suppression, and in the latter case it is argued one may directly access the ``high-energy quanta" of the black hole atmosphere\refs{\LawrenceSG,\Frolov}.  Interestingly, the achievable enhancement of energy loss due to mining matches that of strings, and both are limited by backreaction effects\BrownUN.

These raise the possibility of ``overfilling" a black hole with information, which we define to mean that the entanglement entropy $S(M)$ of the black hole with the radiation exceeds the Bekenstein-Hawking value $S_{BH}(M)$.\foot{ A limiting case of overfilling is formation of a Planck-sized black hole remnant.}  Specifically, consider a black hole that has evaporated to $t_{Page}$, where $S(M)=S({\rm radiation})\sim S_{BH}(M)$.  We could then thread the black hole with strings, or begin mining, and without a corresponding flow of information, $S_{BH}(M)$ would quickly fall below $S(M)$.

One possibility is that the quanta responsible for the enhanced energy flux also accomplish the needed information transfer.  In the string case, this could arise from the nonlocal interactions \hnl\ responsible for the transfer also involving the new Hawking modes on the string.  A priori the story could be the same for mining, although there is a question\refs{\AMPS} whether there is a complete and consistent story compatible with mining.  Perhaps mining indicates deeper inconsistencies in this picture, once its details are properly understood.

Another possible way to address these questions is to return to the generic case of extra energy flux, exceeding that of Hawking, perhaps dictated  by the final consistency of the unitary nonlocal mechanics.  Suppose first that this flux begins after the enhancement from strings/mining, perhaps with some time lag.  The string/mining enhanced flux decreases the black hole lifetime to $t_{decay}$, and to avoid overfilling, we need
\eqn\extraflux{-{dS\over dt} > {S_{BH}(M)\over t_{decay}} > {S_{BH}(M)\over M} {dE_{enhanc}\over dt}\ .}
If $\Ebar$ is the $M$-dependent average energy for the extra quanta carrying the information, the extra flux of energy is 
\eqn\extraM{{dE\over dt} = - \Ebar {dS\over dt} > \Ebar {S_{BH}(M)\over M} {dE_{enhanc}\over dt}\ .}
It is hard to motivate an enhanced flux with $\Ebar\ll T_H$, given the black hole radius $\sim 1/T_H$.  If instead $\Ebar\roughly>T_H$, the extra flux significantly accelerates the black hole decay, again raising the possibility of overfilling.

An alternative is that the dynamics requires information transfer to begin before the Page time, $\tret<t_{Page}$.  Such a ``young" black hole has not yet had time to build up entanglement with Hawking radiation comparable with $S_{BH}$; the typical entanglement with Hawking quanta is given by $S\sim \tret/R$.  For $\tret\ll t_{Page}$, much less entanglement needs to be transferred.  The transfer beginning at $\tret$ can also be characterized by $-dS/dt$ and $\Ebar(M)$.  Even $\Ebar\gg T_H$ doesn't necessarily overfill, if one starts with $S(M)\ll S_{BH}(M)$.  If such early transfer is part of the underlying nonlocal mechanics, this can also be ``nonviolent" to infalling observers ({\it e.g.} as described above), for sufficiently soft $\Ebar$, and given the relatively low needed transfer rates.

Generic models describing necessary information transfer predict extra energy flux from a black hole.  Constraints from thought experiments also suggest a possible need for extra flux.  Constraining consistency arguments are worth further exploration, and may dictate necessary scales.  In particular, if the ``active" atmosphere modes include those to a minimum wavelength $L<R$, there are of order $(L/R)^{D-1}$  degrees of freedom available to carry the information outward, increasing the achievable rate of information transfer at the price of more rapid black hole disintegration.  The scales $L$, $\tret$ could be set, {\it e.g.}, by powers, $L\sim R^p$ with $0<p<1$, $\tret\sim R^q$, constrained by consistency and determined by dynamics.

\newsec{Gauge equivalence, the problem with nice slices, and complementarity}

\slices\ shows three kinds of slices.  In the LQFT framework, we expect each to provide gauge equivalent descriptions of the state.  For example,  to relate a given state on the Schwarzschild slice shown in the figure to that on the nice slice, we could construct an interpolating set of slices, and evolve the state by an operator $U(t^{NS},t^S)$ defined as in \ulqft; in words, the gauge transformation is the integral of the hamiltonian over the region between the slices.  This clearly has action on the state only in the region where the slices don't match, and so is trivial at spatial infinity.  

A similar transformation relating the nice- and natural-slice states requires the completion of the natural-slice degrees of freedom corresponding to core states of the black hole.  The preceding section gave one approach to this, by introducing a cutoff at $r=R_m$, and then describing the states inside this radius in terms of a nice slicing with minimum radius $R_c<R_m$.  Here, the dynamics is frozen, $U(t_2,t_1)\approx 1$, due to vanishing of the lapse.  Or, one could allow the natural slices to enter the strong curvature region, where a spacetime description fails.

While rapid scrambling, {\it e.g} as modeled by \hscram, is not necessarily unexpected for such core states, in a strongly fluctuating regime, the preceding discussion indicates that there are gauge equivalent descriptions where evolution of the state is frozen.  This prediction of LQFT, which is just a statement of the ``nice-slice argument," is at odds with unitary evolution, and apparently cannot be right.  Thus, such a nice slice description apparently must fail, at least after evolution through sufficiently long times.  Ref.~\refs{\QBHB} gave two arguments for such failure, by the time $t_{Page}$ -- the first based on strong backreaction encountered if reference quanta are introduced to give a gauge-invariant description of the state, and the second based on large effects of couplings between fluctuations over long times.\foot{Another argument, based on vanishing of the lapse, was given in \refs{\Arkank}.  However, we see from \ulqft\ that this vanishing simply implies frozen evolution\refs{\QBHB}.}  It should also be noted that comparison of parts of the state at very large separations along the slice also involves enormous relative boosts\refs{\bigboosts}.

The current proposal is that in the regime where the LQFT nice-slice evolution is no longer reliable, it is replaced by dynamics that transfers quantum information to the black hole atmosphere, such as modeled by \hnl, and this dynamics may be accompanied by scrambling described by \hscram.  It is important to ask whether this picture introduces any basic inconsistencies.  Note, in particular, that the picture implies a ``weak" form of complementarity: when quantum information is transmitted from a core state to the atmosphere, it must be ``erased" from the core state, and no longer accessible there, due to the prohibition of cloning of information.  But, this erasure needs only apply to degrees of freedom a time $\sim\tret$ after they fell into the black hole, where the integrity of the nice slice description is suspect.  This suggests the possibility of a consistent description -- given sufficient departure from LQFT.
	
Note that a similar discussion applies to the Schwarzschild description, and could justify a version of strong complementarity, as part of a  gauge-equivalent formulation of dynamics.  Namely, in the Schwarzschild slicing, the lapse vanishes and infalling states freeze at the horizon.  After a time $\sim \tret$ we expect this not to be a good description, and the information should transfer to outgoing radiation.  In this gauge, there is no manifest interior description of the black hole.

We also note that if the correct quantum description is based on networked Hilbert spaces, as suggested in \GiddingsUE, the preceding description of gauge transformations carries over:  the expected form of such a transformation in this more basic framework is a ``local" unitary transformation, which acts differently on different factors of the network, as discussed in \GiddingsUE.

\newsec{Outlook}

If quantum mechanics governs nature, our need for a consistent description of black holes provides a potentially powerful guide to the fundamental quantum physics.  A possible analogy is the classical atom, where inconsistency forced discovery of the principles of quantum mechanics.  

A key question is whether such a more basic quantum description yields objects that still look like black holes, or whether these are replaced by  star-like massive remnants, violently departing from black hole expectations.  Ref.~\AMPS\ has argued for the latter, in the form of a ``firewall," and gave arguments, elaborated, {\it e.g.} in \refs{\Lenny,\Raphael}, that postulated black hole complementarity  in fact implies this scenario.
	
Nonlocality with respect to the semiclassical geometrical picture appears necessary to save quantum mechanics, but a very different scenario, with less violent nonlocality, is proposed to describe black hole mechanics.  Here gross features of black holes may be preserved, and in particular the possibility of at least some infalling observers probing black hole interiors without encountering extreme violence near the would-be horizon. 
However, some black hole expectations may be modified; sufficiently sensitive measurements by infalling observers may detect extra quanta, and generic such evolution predicts more rapid black hole disintegration than that of Hawking.  A critical question is whether there is a framework that is not inconsistent, either internally, or with basic principles like quantum mechanics and Lorentz invariance,  and which embeds local quantum field theory and basic features of gravity as an approximation.

This paper has described ``phenomenological" simplified models for such dynamics.  At first sight, these are nothing more than toy models, or effective parameterizations of possible collective dynamics of a black hole.  However, stringent constraints of consistency suggest that these models may yield more.  We recall that attempting to consistently parameterize quantum dynamics that is Lorentz invariant and local leads to the full structure of quantum field theory.  The nontriviality of the constraints on the present structure are exemplified by the tension between quantum information transfer and classical spacetime structure\refs{\Mathurbit,\SBGmodels,\GiddingsUE,\AMPS}.  The tightness of these  constraints on the more basic framework --  networked Hilbert spaces or otherwise -- may largely determine its nature.	

\bigskip\bigskip\centerline{{\bf Acknowledgments}}\nobreak

I thank J. Hartle, D. Marolf, J. Polchinski, and Y. Shi for helpful conversations.  This work  was supported in part by the Department of Energy under Contract DE-FG02-91ER40618 and by a Simons Foundation Fellowship, 229624, to Steven Giddings.

\appendix{A}{Qubit models for core states and scrambling/transfer}

A simple toy model for black hole core states is as a string of $N$ qubits.  One might expect $N\sim S_{BH}$, although as section 2.3 discusses, the number of internal states might be constrained to be smaller.  Consider $N$ ladder operators obeying (anti)commutation relations
\eqn\anticomm{\{b_i,b_i\}=\{b_i^\dagger,b_i^\dagger\}=0\ ,\quad \{b_i,b_i^\dagger\}=1\ ,\quad  [b_i,b_j]=[b_i^\dagger,b_j]=0\ {\rm  for}\  i\neq j\ ,}
differing from usual fermionic operators by a Klein transformation.  These operators excite qubits:  $b_i^\dagger |0\rangle_i = |1\rangle_i$.  
Then nonlocal transfer can be modeled as in  \hnl, and scrambling as in \hscram, so the departure from the ``frozen" evolution of nice slices in LQFT is
\eqn\hnew{ H'= {1\over R} \sum_{i,a} \calN_{{\bar a}i}(t) a^\dagger_a b_i + {1\over R} \sum_{i,j} {\cal S}_{{\bar \imath }j} b_i^\dagger b _j + {\rm h.c.}\ .}
Interactions with incoming matter can likewise be modeled as in \Hexp.  Care is needed to account for all incoming information, but this serves as a simplified model to explore dynamics of scrambling and transfer.  We leave its study for future work.

\listrefs
\end